# Day-ahead Operational Planning with Enhanced Flexible Ramping Product: Design and Analysis

Mohammad Ghaljehei, *Student Member, IEEE,* Mojdeh Khorsand, *Member, IEEE*

*Abstract*— New resource mix, e.g., renewable resources, are imposing operational complexities to modern power systems by intensifying uncertainty and variability in the system net load. This issue has motivated independent system operators (ISOs), e.g., California ISO (CAISO), to add the flexible ramping product (FRP) to their day-ahead (DA) market models. Such structural changes in the DA market formulation require further analyses and detailed design to ensure adequate operational flexibility, market efficiency, and reliability. This paper conducts a comprehensive study to: (a) augment existing DA market models with enhanced FRP design in order to schedule ramp capabilities that are more adaptive with respect to the real-time (RT) condition, and (b) design corresponding market payment policies that accurately reflect the value of the added flexibility through enhanced FRP design. The proposed FRP design can be implemented in present-day system operations with minimal disruption to existing DA market models. Performance of the proposed DA market model, which includes the enhanced FRP design, is compared against the DA market model with existing FRP design through a validation methodology based on RT unit commitment model. This validation methodology mimics fifteen-minute market of CAISO. The proposed method is tested on an IEEE 118-bus test system.

## NOMENCLATURE

*Sets and Indices*

| | |
|---|---|
| $g$ | Index of generators, $g \in G$. |
| $g(n)$ | Set of generators connected to node $n$. |
| $k$ | Index of transmission lines, $k \in K$. |
| $n$ | Index for buses, $n \in N$. |
| $t$ | Index for time periods, $t \in T$. |

*Parameters and Constants*

| | |
|---|---|
| $c_g^{NL}, c_g^{SD}, c_g^{SU}$ | No-load, shutdown and startup costs of unit $g$. |
| $P_g^{max}, P_g^{min}$ | Maximum output and minimum output of unit $g$. |
| $R_g^{HR}, R_g^{15}$ | Hourly and 15-min ramp rates of unit $g$. |
| $R_g^{SU}, R_g^{SD}$ | Startup and shutdown ramp rate of unit $g$. |
| $UT_g, DT_g$ | Minimum up time and down time of unit $g$. |
| $FRup_t$ | Hourly ramping up requirement in period $t$. |
| $FRdown_t$ | Hourly ramping down requirement in period $t$. |
| $FRup_{t,(.)}^{ih}$ | 15-min ramping up requirement in period $t$. |
| $FRdown_{t,(.)}^{ih}$ | 15-min ramping down requirement in period $t$. |
| $NL_t, NL_{t,(.)}^{ih}$ | Hourly and 15-min system net load in period $t$. |
| $NL_t^{max}$ | Maximum hourly system net load in period $t$. |
| $NL_{t,(.)}^{ih,max}$ | Maximum 15-min system net load in period $t$. |
| $NL_t^{min}$ | Minimum hourly system net load in period $t$. |
| $NL_{t,(.)}^{ih,min}$ | Minimum 15-min system net load in period $t$. |
| $P_k^{max}$ | Thermal rating of transmission line $k$. |
| $c_g^p$ | Variable cost of unit $g$ ($/MWh). |
| $Load_{nt}$ | Load at bus $n$ during time period $t$. |
| $PTDF_{nk}$ | Power transfer distribution factor for line $k$ for an injection at bus $n$. |

*Variables*

| | |
|---|---|
| $P_{gt}$ | Output of unit $g$ at period $t$. |
| $P_{nt}^{inj}$ | Net power injection at bus $n$ at period $t$. |
| $u_{gt}, v_{gt}, w_{gt}$ | Unit commitment, startup, and shutdown variables for unit $g$ in period $t$. |
| $ur_{gt}, dr_{gt}$ | Hourly up and down FRP award of unit $g$ in period $t$. |
| $ur_{gt}^{ih}, dr_{gt}^{ih}$ | Intra-hour ramping up and down auxiliary variables of unit $g$ in period $t$. |

## I. INTRODUCTION

The recent rapid integration of variable energy resources, such as wind and solar, causes new operational challenges for power systems. For example, 26 percent of total generation of CAISO in 2018 was served by non-hydro renewable, an increase from 24, 22, and 18 percent in 2017, 2016, and 2015, respectively [1]. One emerging challenge, due to this trending increase, is the growth of variability and uncertainty in the net load (i.e., actual system load minus the scheduled interchange and total renewable generation). The recent trend of outages and blackouts, such as California rolling outages, also raises concerns about generation capacity shortage and the flexibility needs [2]-[3]. In such situations, if there is insufficient ramp capability in the system, the ISOs may not be able to serve the demand totally; there may not be enough flexible generation resources in the system to follow the net load changes. Moreover, such generation shortage can cause high penalty prices during the RT market processes and consequently create market inefficiency in the long run [4]-[5]. In some markets, including the one in CAISO, there has been acknowledgement that traditional regulation and contingency-based reserve products are not able to provide sufficient ramp capability; therefore new market products are necessary [6]. As a result, some ISOs, e.g., CAISO [7] and midcontinent ISO (MISO) [8], have been augmenting their market models with ramping products in order to meet the net load variability and uncertainty and attain higher responsiveness from existing flexible resources. In this paper, such products are referred to as FRP, although they are also called "flexiramp" in CAISO [7], and "ramp capability" in MISO [8].

There have been intentions of modeling and procuring FRP in the DA market in order to ensure resource adequacy for meeting the ramping requirements in RT. For example, CAISO has started market initiatives to add FRP to its DA market to correctly position and commit resources to address uncertainty and variability in the net load previously left to its RT market [7] and [9]. In previous work and existing industry proposals [10]–[13] and [7], the DA FRP is designed in such a way that resultant added ramp capabilities must be able to respond to foreseen and unforeseen changes in net load between hour $t$ and next hour, $t + 1$. However, this FRP design (called general FRP design in this paper) does not consider magnitude of intra-hour net load changes in the following market processes which have shorter scheduling granularity (e.g., 15-min and 5-min intervals). Essential goal of modeling the hourly FRP is to increase the system ramp capability to follow the realized net load in next RT market processes [7]. In this situation, the general FRP may not be able to accommodate the steep realized RT net load changes as they happen in the shorter periods of time and their effects are not considered when the hourly FRP decisions are made in DA. Two examples of such cases are shown in Cases I and II in Fig. 1. With the above discussion, there is a need for detailed design of the FRP in the DA market in order to ensure that enough ramp capabilities are available from flexible and yet non-expensive units; an effective FRP modeling in DA operation prepositions and commits the generation units and enables them to respond to sharp transitions of RT net load. CAISO also acknowledges the necessity of meeting 15-min ramping needs by hourly reserved ramp capabilities and DA schedules as specified in [9] and [14]: "*Steep net load differences between 15-minute intervals (granularity differences) may result in 15-minute ramp infeasibility due to mid-point to mid-point hourly scheduling*".

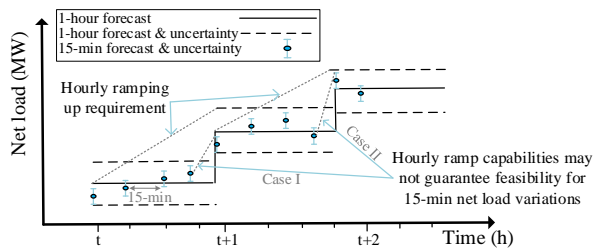

Fig. 1. Hourly FRP versus 15-min variability and uncertainty.

Although it is vital to evaluate both the efficiency and reliability, it is also necessary to evaluate the incentive structure for the ramp-responsive resources. It is pertinent to note that when the energy and ancillary service products are co-optimized, and no bids are submitted for the ancillary service product, the price for ancillary service is based on the lost opportunity cost [15]. Similarly, in the context of FRP, as a new emerging ancillary service, the flexible resources are asked to perform in different ways for providing the ramp capability while no bids/offers are submitted [5]. Thus, proper incentives and/or compensations for the lost opportunity cost should be provided in order to encourage following market decisions; however, this step often is ignored in the majority of studies. The goal of this paper is to complement proposed efforts by industry and previous work to more precisely schedule DA FRP with minimal added computational burdens while appropriately incentivizing flexible resources to provide FRP.

*A. Literature review*

A number of prior work [4]–[6], [15]–[17] have been focused on the FRP in RT markets since the main purpose of the FRP is to improve the ramp capabilities in RT. There also have been intentions in some work [10]–[13] toward implementing and procuring some of the ramp capabilities in the DA market. Reference [10] has proposed an optimization model for energy storage aggregators to maximize its profit by biding for and procuring FRP in the DA energy and reserve markets. However, it is pertinent to note that the FRP is considered as a not-biddable product [5]. A wind power ramping product (WPRP) has been proposed by [11] to allow wind resources to participate in the FRP market. The WPRP is designed for responding to the needed ramping requirements in order to ensure sufficient ramp capabilities in the RT operations. However, due to their relative low operational costs, wind units may not be good options to procure the ramp capability [18]. Reference [12] has proposed an integrated stochastic DA scheduling model to allocate FRP for managing the variability and uncertainty of the renewable energy system. A non-deterministic FRP design is proposed in [13] to adequately allocate the ramp capacities in the DA market. The proposed model, which is an affinely adjustable robust UC model, is based on a cost-free ramp capacity procurement procedure. In spite of appealing results, [12] and [13] have respectively utilized stochastic programming and robust optimization methods, both of which are under the umbrella of advanced stochastic programming techniques [19]. However, due to scalability issues, the FRP was originally designed to keep the market models close to the existing practice while addressing variability and uncertainty [6]. Furthermore, in compliance with common practice, proper market-based incentive policies corresponding to the modified FRP designs should be developed for the flexible resources that participate in FRP market; however, this step is ignored in [10]–[13]. Overall, such structural changes, i.e., inclusion of FRP in DA market formulation, requires detailed analyses and design in order to ensure adequate operational flexibility, market efficiency, and pricing. In the same direction of proposal of CAISO [7], the focus of this paper will be on the efficient design of DA FRP and associated incentivizing policies.

This paper first presents the formulation of a DA market model with general FRP constraints. Then, it sheds light on a subtle issue that can potentially happen in the next market processes after DA market, as a result of procurement of DA ramp capabilities only based on the hourly ramping requirements. Then, a new FRP design is proposed to address this issue. In the proposed formulation, the DA FRP design is modified to satisfy the 15-min net load variability and uncertainty while scheduling hourly FRPs in the DA market. The proposed approach overcomes the concerns raised by CAISO in [9] and [14] regarding the 15-min ramp infeasibility. New incentivizing policies are driven for the proposed FRP design in order to properly encourage the flexible resources that provide the enhanced ramp capability. Finally, to effectively evaluate different FRP designs from reliability and market efficiency points of view, a validation methodology is proposed that is similar to the RT unit commitment (RTUC) processes of the CAISO. The main contributions of this paper are:
- Propose a novel FRP requirement in DA market in order to preposition and commit flexible resources to respond to sharp transitions in the RT net load.
- Design corresponding market payment policies using primal-dual market formulations that accurately reflect the value of awarded ramp capabilities through the proposed FRP design.
- Create a validation methodology that mimics the RT market of CAISO in order to evaluate both efficiency and reliability of the proposed FRP design against general FRP design.
- Conduct a comprehensive evaluation of the proposed FRP design, including its impacts on the market efficiency, settlements and system reliability.

The rest of the paper is organized as follows. Section II presents model formulation including the enhanced FRP design. Section III focuses on market payment mechanism for the proposed FRP design. Section IV presents the validation phase methodology. Finally, simulation results are discussed in Section V while section VI concludes the paper.

## II. MODEL FORMULATION

The general FRP design within the DA market model is formulated in Section II.A, which is based on [10], [11] and [7]. Then, a subtle issue that can potentially happen in the next market processes after DA market with the shorter granularity or time resolution is discussed in Section II.B. This subtle issue is the motivation behind the proposed DA FRP design presented in Section II.C to improve the general FRP design from reliability and efficiency points of view.

### A. DA market model with general FRP design

DA market model with general FRP design and requirement is presented in (1a)-(1w). The objective function (1a) is to minimize total operating costs (i.e., variable operating costs, no-load costs, startup costs, and shutdown costs):

$$\text{minimize} \sum_g \sum_t (c_g^p P_{gt} + c_g^{NL} u_{gt} + c_g^{SU} v_{gt} + c_g^{SD} w_{gt}) \quad (1a)$$

The set of constraints are shown in (1b)-(1w). Minimum up and down time constraints of the generators are enforced by (1b) and (1c), while constraints (1d) and (1e) ensure hourly ramp rate limits. Constraints (1f)-(1i) ensure system-wide and nodal power balance, calculate power flow, and impose transmission line flow limits. Constraints (1j) and (1k) model the relationship between the commitment variable, the startup variable, and shutdown variable. Constrains (1l)-(1o) model the binary commitment decision ($u_{gt}$) and the startup ($v_{gt}$) and shutdown ($w_{gt}$) decisions, respectively.

$$\sum_{s=t-UT_g+1}^{t} v_{gs} \leq u_{gt}, \forall g, t \in \{UT_g,..,T\} \quad (1b)$$
$$\sum_{s=t-DT_g+1}^{t} w_{gs} \leq 1 - u_{gt}, \forall g, t \in \{DT_g,..,T\} \quad (1c)$$
$$P_{gt} - P_{gt-1} \leq R_g^{HR} u_{gt-1} + R_g^{SU} v_{gt}, \forall g, t \geq 2 \quad (\gamma_{gt}^-) \quad (1d)$$
$$P_{gt-1} - P_{gt} \leq R_g^{HR} u_{gt} + R_g^{SD} w_{gt}, \forall g, t \geq 2 \quad (\gamma_{gt}^+) \quad (1e)$$
$$\sum_{g \in g(n)} P_{gt} - Load_{nt} = P_{nt}^{inj}, \forall n, t \quad (\delta_{nt}) \quad (1f)$$
$$\sum_n P_{nt}^{inj} = 0, \forall t \quad (\lambda_t) \quad (1g)$$
$$\sum_n P_{nt}^{inj} PTDF_{nk} \leq P_k^{max}, \forall k, t \quad (F_{kt}^+) \quad (1h)$$
$$-P_k^{max} \leq \sum_n P_{nt}^{inj} PTDF_{nk}, \forall k, t \quad (F_{kt}^-) \quad (1i)$$
$$v_{gt} - w_{gt} = u_{gt} - u_{g,t-1}, \forall g, t \quad (1j)$$
$$v_{gt} + w_{gt} \leq 1, \forall g, t \quad (1k)$$
$$0 \leq v_{gt} \leq 1, \forall g, t \quad (1l)$$
$$0 \leq w_{gt} \leq 1, \forall g, t \quad (1m)$$
$$u_{gt} \in \{0,1\}, \forall g, t \quad (1o)$$

Constraints (1p)-(1w) are associated to general FRP constraints. The generator output limits including up and down FRPs are presented by (1p) and (1q). Constraints (1r) and (1s) limits the ramp up and down capabilities to the hourly generators ramp rate, respectively. The hourly FRP requirements are met via (1t)-(1u). Finally, the hourly up and down FRP requirements are calculated through (1v) and (1w).

$$p_{gt} + ur_{gt} \leq P_g^{max} u_{gt}, \forall g, t \quad (\alpha_{gt}^+) \quad (1p)$$
$$p_{gt} - dr_{gt} \geq P_g^{min} u_{gt}, \forall g, t \quad (\alpha_{gt}^-) \quad (1q)$$
$$ur_{gt} \leq R_g^{HR} u_{gt}, \forall g, t \quad (\beta_{gt}^+) \quad (1r)$$
$$dr_{gt} \leq R_g^{HR} u_{gt}, \forall g, t \quad (\beta_{gt}^-) \quad (1s)$$
$$\sum_g ur_{gt} \geq FRup_t, \forall t \quad (\pi_t^+) \quad (1t)$$
$$\sum_g dr_{gt} \geq FRdown_t, \forall t \quad (\pi_t^-) \quad (1u)$$
$$FRup_t = max\{NL_{t+1}^{max} - NL_t, 0\}, \forall t \leq 23 \quad (1v)$$
$$FRdown_t = max\{NL_t - NL_{t+1}^{min}, 0\}, \forall t \leq 23 \quad (1w)$$

In the above formulation, the dual variable related to each primal constraint is referenced after the primal constraint. Furthermore, in order to investigate the impact of FRP formulation and requirements, other ancillary services are excluded from the model for the sake of clarity. However, the analyses and results are generalizable to other market models where various ancillary services requirements are included.

### B. DA resource adequacy to meet RT ramping needs with general FRP design

In the market model with the general FRP design presented by (1a)-(1w), the FRP formulation is designed in such a way that resultant added ramp capabilities be able to respond to foreseen (variability) and unforeseen (uncertainty) changes in the net load between the hour $t$ and the specified target hour $t + 1$. Although the procured DA FRP is aimed to increase the ramp capabilities to meet the steep ramping needs in the next RT markets such as the fifteen-minute market (FMM) [7], this design disregards magnitude of the intra-hour net load changes. The FMM seeks to meet the balance between supply and net load in 15-min intervals. The net load variation between the 15-min intervals can potentially experience steeper slope, which happen in a shorter time than the hourly net load variations – see case I and case II in Fig. 1. This issue, i.e., steep net load differences between 15-minute intervals and potential 15-min ramp infeasibility due to mid-point to mid-point hourly scheduling, is also raised by the CAISO [9] and [14]. Thus, (1a)-(1w) may not ensure adequate procurement of up and down FRPs to cover potential steep net load changes in FMM, i.e., the next closest RT market process to DA market.

### C. Feasibility of hourly DA FRP awards against intra-hour 15-min variability and uncertainty

In this paper, the concept of intra-hour ramping requirement is incorporated into the DA FRP design, and additional FRP

constraints are designed to ensure DA ramp capabilities are adequate for accommodating 15-min net load variability and uncertainty. The 15-min ramping requirements should accommodate: (i) foreseen variability in the 15-min net load, and (ii) unexpected net load variations between two successive time intervals considering a desired confidence level. Thus, intra-hour ramp up requirements are formulated via (2a)-(2d).

$$FRup_{t,0min}^{ih} = max\{NL_{t,15min}^{ih,max} - NL_{t,0min}^{ih}, 0\}, \forall t \quad (2a)$$
$$FRup_{t,15min}^{ih} = max\{NL_{t,30min}^{ih,max} - NL_{t,15min}^{ih}, 0\}, \forall t \quad (2b)$$
$$FRup_{t,30min}^{ih} = max\{NL_{t,45min}^{ih,max} - NL_{t,30min}^{ih}, 0\}, \forall t \quad (2c)$$
$$FRup_{t,45min}^{ih} = max\{NL_{t+1,0min}^{ih,max} - NL_{t,45min}^{ih}, 0\}, \forall t \quad (2d)$$

(2a)-(2c) represent 15-min ramp up requirements for each two successive 15-min intervals of hour $t$. (2d) shows the ramp up requirement associated to the last 15-min interval of hour $t$ and first 15-min interval of hour $t+1$ (see Fig. 1 for clarity). The enhanced formulation uses auxiliary variables, i.e., $ur_{gt}^{ih}$, for intra-hour ramping schedules from resources. Summation of these auxiliary variables should satisfy all intra-hour 15 min ramping requirements as shown in (2e).

$$\sum_{\forall g} ur_{gt}^{ih} \geq max(FRup_{t,0min}^{ih}, FRup_{t,15min}^{ih},$$
$$FRup_{t,30min}^{ih}, FRup_{t,45min}^{ih}), \forall t \quad (\pi_t^{ih,+}) \quad (2e)$$

where $ur_{g,t}^{ih}$ is limited to 15-min ramp rate.

$$ur_{g,t}^{ih} \leq R_g^{15} u_{g,t}, \forall g, t \quad (\beta_{gt}^{ih,+}) \quad (2f)$$

Finally, in order to ensure sufficiency of DA FRP to accommodate 15-min variability and uncertainty, constraint (2g) is included. This constraint identifies that DA FRP of each ramp-responsive resource should be greater than its corresponding intra-hour ramping schedule.

$$ur_{g,t}^{ih} \leq ur_{g,t}, \forall g, t \quad (\omega_{gt}^{+}) \quad (2g)$$

Similar formulation can be proposed for the enhanced DA downward FRP design, which are presented in (2h)-(2n):

$$FRdown_{t,0min}^{ih} = max\{NL_{t,0min}^{ih} - NL_{t,15min}^{ih,min}, 0\}, \forall t \quad (2h)$$
$$FRdown_{t,15min}^{ih} = max\{NL_{t,15min}^{ih} - NL_{t,30min}^{ih,min}, 0\}, \forall t \quad (2i)$$
$$FRdown_{t,30min}^{ih} = max\{NL_{t,30min}^{ih} - NL_{t,45min}^{ih,min}, 0\}, \forall t \quad (2j)$$
$$FRdown_{t,45min}^{ih} = max\{NL_{t,45min}^{ih} - NL_{t+1,0min}^{ih,min}, 0\}, \forall t \quad (2k)$$
$$\sum_{\forall g} dr_{gt}^{ih} \geq max(FRdown_{t,0min}^{ih}, FRdown_{t,15min}^{ih},$$
$$FRdown_{t,30min}^{ih}, FRdown_{t,45min}^{ih}), \forall t \quad (\pi_t^{ih,-}) \quad (2l)$$
$$dr_{g,t}^{ih} \leq R_g^{15} u_{g,t}, \forall g, t \quad (\beta_{gt}^{ih,-}) \quad (2m)$$
$$dr_{g,t}^{ih} \leq dr_{g,t}, \forall g, t \quad (\omega_{gt}^{-}) \quad (2n)$$

Finally, the proposed DA market model with the enhanced FRP requirements is shown in (2o)-(2p).

$$minimize \sum_g \sum_t (c_g^p P_{gt} + c_g^{NL} u_{gt} + c_g^{SU} v_{gt} + c_g^{SD} w_{gt}) \quad (2o)$$

subject to:

(1b)-(1w) and (2a)-(2n) (2p)

Note that, although with the proposed FRP design, the DA solution will accommodate the 15-min net load changes, the DA scheduling framework will still be kept based on the hourly granularity to be consistent with current industry practice.

III. MARKET PAYMENT MECHANISM FOR THE PROPOSED FRP DESIGN

Generally, the generators do not give bids for ramping products, instead the shadow price (dual variable) of the ramping requirements is employed to calculate the payment for FRP awards [5], [6] and [17]. This payment is to compensate ramp qualified generators in terms of an opportunity cost and is composed of a capacity component and a ramping component [17]. With the new FRP design for DA market model, (2o)-(2p), a new market payment mechanism is essential for generators that provide the enhanced ramp capability product to properly incentivize them to follow the ISO signal. In this paper, the primal-dual formulation and complementary slackness (CS) conditions are leveraged to derive corresponding market payment mechanism for the proposed FRP design.

The DA market model (2o)-(2p) is a mixed-integer linear programing (MILP) and therefore the dual formulation is not well-defined. To derivate of the dual variables and dual formulation, the linear programming problem created in the node with best feasible integer solution in the Branch-and-Bound algorithm can be utilized. After deriving full dual formulation, dual constraints related to variables $ur_{g,t}$ and $ur_{g,t}^{ih}$ can be written as (3a) and (3b), respectively.

$$-\alpha_{gt}^+ - \beta_{gt}^+ + \pi_t^+ + \omega_{gt}^+ \leq 0, \forall g, t \quad (ur_{g,t}) \quad (3a)$$
$$\pi_t^{ih,+} - \beta_{gt}^{ih,+} - \omega_{gt}^+ \leq 0, \forall g, t \quad (ur_{g,t}^{ih}) \quad (3b)$$

For calculating the new market payment for a generator with up DA FRP award, CS condition associated with constraints (3b) and (2g) are respectively implemented as follows.

$$\omega_{gt}^+ ur_{g,t}^{ih} = \pi_t^{ih,+} ur_{g,t}^{ih} - \beta_{gt}^{ih,+} ur_{g,t}^{ih}, \forall g, t \quad (3c)$$
$$\omega_{gt}^+ ur_{g,t}^{ih} = \omega_{gt}^+ ur_{g,t}, \forall g, t \quad (3d)$$

Then, substitution of (3d) in (3c) results in (3e).

$$\omega_{gt}^+ ur_{g,t} = \pi_t^{ih,+} ur_{g,t}^{ih} - \beta_{gt}^{ih,+} ur_{g,t}^{ih}, \forall g, t \quad (3e)$$

Furthermore, CS condition associated with the constraint (3a) can be written as (3f).

$$\omega_{gt}^+ ur_{g,t} = \alpha_{gt}^+ ur_{g,t} + \beta_{gt}^+ ur_{g,t} - \pi_t^+ ur_{g,t}, \forall g, t \quad (3f)$$

Finally, by substituting (3f) in (3e) and performing rearrangements, equation (3g) can be derived.

$$\pi_t^+ ur_{g,t} + \pi_t^{ih,+} ur_{g,t}^{ih} = \alpha_{gt}^+ ur_{g,t} + \beta_{gt}^+ ur_{g,t} + \beta_{gt}^{ih,+} ur_{g,t}^{ih} \quad (3g)$$

In (3g), right-hand side terms, i.e., $\alpha_{gt}^+ ur_{g,t}, \beta_{gt}^+ ur_{g,t}$, and $\beta_{gt}^{ih,+} ur_{g,t}^{ih}$, are respectively associated with the opportunity cost for the generator in terms of withholding some portion of its capacity and ramp capabilities. Based on the equation (3g), left-hand side, i.e., $\pi_t^+ ur_{g,t} + \pi_t^{ih,+} ur_{g,t}^{ih}$, is the flexible ramp up payment of generator $g$ at time interval $t$ for procuring enhanced upward FRP. Hence, a generator that is ramp up qualified should be compensated based on $OC_g^{Rup,new}$ for its up FRP provision for the entire scheduling horizon.

$$OC_g^{Rup,new} = \sum_t (\pi_t^+ ur_{g,t} + \pi_t^{ih,+} ur_{g,t}^{ih}), \forall g \quad (3h)$$

Similarly, it can be derived that the ramp down qualified generator for procuring down DA FRP should be compensated based on $OC_g^{Rdown,new}$ for the entire scheduling horizon.

$$OC_g^{Rdown,new} = \sum_t (\pi_t^- dr_{g,t} + \pi_t^{ih,-} dr_{g,t}^{ih}), \forall g \quad (3i)$$

IV. VALIDATION METHODOLOGY

A RT validation methodology is needed to compare the performance of the proposed FRP design and the general FRP design. Some of the previous work have implemented RT economic dispatch performed every 5-minute as the validation methodology while ignoring the other market processes, i.e., the FMM, that happen between the DA market and RT economic dispatch market. The FMM is the first market which (i) makes the preliminary modifications to the posted DA market solutions for meeting RT load, and (ii) experiences the impacts of DA market reformulations and enhancements. Hence, it can be a good candidate to evaluate the performance of the DA market decisions from the flexibility, economic and reliability points of view. This paper creates a RT validation methodology, which mimics FMM of CAISO. This methodology is used to conduct comprehensive reliability and economic comparisons of proposed method with general FRP model under 15-min net load scenarios. Flowchart of proposed simulation procedure is presented in Fig. 2.

### A. FMM of CAISO [20]

The FMM of CAISO includes four RTUC runs for each trading hour. RTUC is a market process for committing fast-start (FS) units at 15-minute intervals, which is performed on a rolling-forward basis. The four RTUC processes, i.e., RTUC#1, RTUC#2, RTUC#3, and RTUC#4, include time horizons of 60-105 minutes spanning from the previous trading hour and the current trading hour. The binding interval is the second interval of the RTUC run horizon, and the rest are advisory intervals.

### B. Proposed validation phase

For the sake of simplicity in the proposed validation methodology, without loss of generality, all the four RTUC runs are combined into a one-process RTUC run for each trading hour. The one-process RTUC run has the same number of intervals as RTUC#1 (i.e., seven 15-min intervals) and is performed in the same time schedule as the RTUC#1 is performed. It approximately starts 7.5 minutes prior to the first trading hour for $T-45$ minutes to $T+60$ minutes, where $T$ is the top of the trading hour as shown in Fig. 3, and it continues on a rolling-forward basis for the next trading hours. Since each original RTUC process has 1 binding interval, the one-process RTUC simulation would have four successive binding intervals out of 7 intervals. It is pertinent to note that the binding intervals of trading hour $T$ that are also a part of intervals of the next trading hour $T+1$, are used and kept the same (dispatch and commitment) in one-process RTUC run of trading hour $T+1$.

After performing all one-process RTUC runs, the final schedule for the 96 intervals of a day are achieved through putting together all the binding intervals. Furthermore, in each one-process RTUC run, power balance violations are allowed to occur if there are insufficient ramp capabilities to follow the sudden net load changes.

### C. Data transferring from DA market to FMMs

In the proposed validation methodology, the DA market solutions are transferred and embedded into the FMMs for running the multiple one-process RTUC problems. The commitments obtained from DA market model are kept fixed for long-start units in the FMMS. Furthermore, the dispatch modification of the committed units in the DA market model is limited to their 15-min ramp rate in the first interval of RTUC process so as to not deviate much from DA market dispatch decisions. Furthermore, FS units can be further committed to follow the realized net load if the ramping shortage occurs.

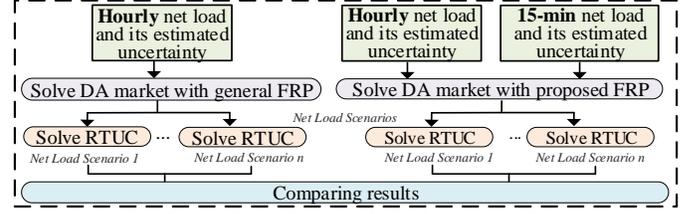

Fig. 2: Flowchart of proposed simulation procedure.

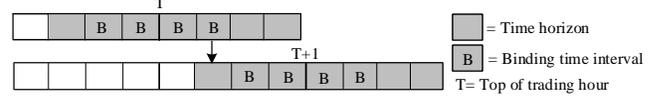

Fig. 3: One-process RTUC run in the validation phase.

## V. NUMERICAL STUDIES AND DISCUSSION

A 118-bus IEEE test system is employed for testing the proposed framework. CPLEX v12.8 is utilized to solve the MILP models on a computer with an Intel Core i7 CPU @ 2.20 GHz, 16 GB RAM, and 64-bit operating system.

### A. System Data and assumptions

The 118-bus IEEE test system has 54 generators, 186 lines, and 91 loads [21]. The hourly and 15-min net load profiles were adopted from the real data of CAISO [22], as depicted in Fig 4. Simulation results are performed for two different days, Feb 20, 2020 (i.e., first day) and January 20, 2020 (i.e., second day). From Fig. 4, it is clear that the first day is prone to have more severe net load changes compared to the second day. The uncertainty in the net load of CAISO is mainly introduced by renewable resources, especially solar and wind [22]. The error of the hourly net load forecast is assumed to have a Gaussian distribution with zero mean and ~5% standard deviation. Also, 95% confidence level, i.e., 1.96 standard deviations, is considered for hourly ramping requirements, i.e., (1v) and (1w). Based on the total probability theory [16] and [23], the relationship between standard deviation of error of hourly and 15-min net load can be quantified, i.e., $\sigma_{hourly} = 2\sigma_{15-min}$. By using this formula, the 15-min ramping requirements formulated by (2a)-(2d) and (2h)-(2k) can be calculated. The confidence level for 15-min ramping requirements is also 95%.

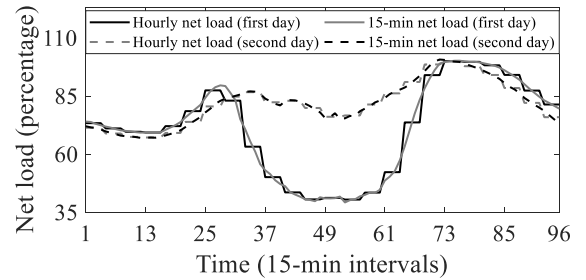

Fig. 4: Hourly net load versus 15-min net load.

In the validation methodology, 18 gas and oil generators, which have the maximum capacity up to 50 MW are considered as the FS units. The violation is allowed to occur if the available ramp capabilities are not enough to follow the realized net load. The penalty price is in term of value of lost load (VOLL), which is chosen as $10000/MWh. Additionally, 500 different 15-min net load scenarios are generated based on the 15-min net load uncertainty for this validation phase.

*B. Proposed FRP design versus general FRP design*

Fig. 5 compares increased number of 15-min commitments of FS units against the total violation for the net load test scenarios in FMMs obtained from the validation phase. The increased number of FS units 15-min commitments represents the increase in the total commitment of FS units from DA market to RT market for total of 96 time intervals. Furthermore, Table I presents the number of scenarios where the proposed approach outperforms the general approach from economic and reliability metrics. According to Fig. 5 and Table I, it can be seen that for almost all the scenarios, the proposed approach provides pareto optimal solutions (with respect to increased number of 15-min commitments of FS units and the violation) compared to the general approach, wherein less violation and increased commitment number occur in the FMMs. The reason is that the proposed approach preemptively takes into account the impacts of 15-min net load variability and uncertainty on the DA FRP decisions. Please note that lower number for commitment increase is an indicative of less discrepancy between the DA and RT markets, which potentially can lead to the reduction of the necessity for expensive adjustments in the RT market processes. These results show the efficiency of the proposed FRP design in quantifying more adaptive FRP requirements in the DA market with respect to the RT condition. Owing to discounting the impacts of the nodal FRP deployment on the physical network limitations (e.g., congestion) in the DA market model, the proposed approach is not expected to fully remove additional need for commitment of FS units in the FMM. That is why in the Fig. 5, the FS units are still required to be turned on to follow the net load. Future work should investigate how to address the deliverability issue associated with the post-deployment of FRP in the RT markets. Fig. 6 compares the real-time operating costs of FMMs (excluding the violation cost with VOLL) against the total violation for the corresponding FRP designs over the net load test scenarios. In Fig. 6, in order to assess the performance of proposed approach under different bids of FS units in RT, the results are presented not only for the FMMs in which the FS units have the same bids as their bids in the DA market, but also for the FMMs with FS units having increased bid (by 15% of the DA bids). It is worth mentioning that if the operating costs include the violation cost (similar to what is usually done in the literature [15], [16], [18], [24]), the comparisons between the operating costs of different models may be subjective as the results are sensitive to the choice of VOLL. In this paper, the real-time operating costs are compared through different approaches so that the comparisons can be more objectively conducted. These approaches include: (i) performing sensitivity analysis and (ii) removing the cost associated with violations and VOLL from the operating costs while comparing the violation as another metric. Based on Fig. 6 and Table I, it can be observed that the proposed approach is effectively capable of reducing the violation while resulting in lower or comparable real-time operating costs (excluding the violation cost with VOLL) to those of the general approach. In other words, considerably less load shedding has been observed for the proposed approach (proposed approach has less or equal violation in 98.4% of scenarios), while in 62.4% and 69.2% of scenarios, the obtained real-time operating costs (excluding the violation cost with VOLL) of proposed approach are lower than those of the general approach for the original and the increased bids, respectively. These results prove the enhanced performance of the proposed FRP design for improving the RT market efficiency from the economic and reliability aspects for different bids in RT market.

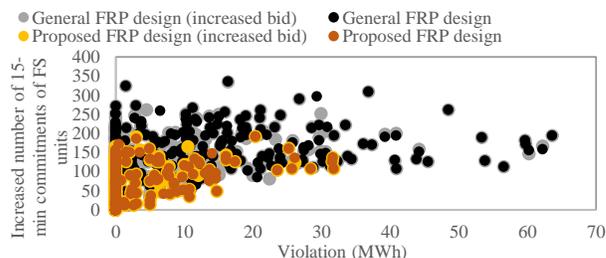

Fig. 5. Increased number of 15-min commitment of the FS units versus violation in RT operation (first test day)

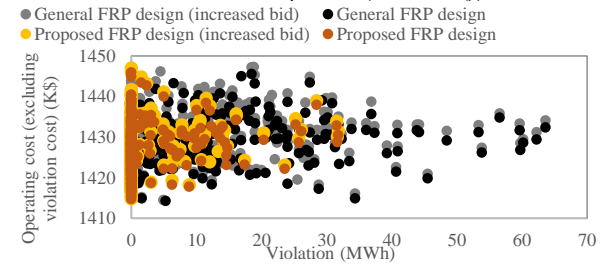

Fig. 6. RTUC operating cost versus violation in RT operation (first test day)

TABLE I. NUMBER OF SCENARIOS WITH IMPROVEMENT OVER ALL TIME INTERVALS IN RT OPERATION (TOTAL NUMBER OF SCENARIOS = 500)

| Metric | |
|---|---|
| # Scenarios with total violation improvement* | 492 |
| # Scenarios with reduction in total commitment of FS units | 478 |
| # Scenarios with cost (excluding violation cost) improvement | 312 |

*Scenarios with same or less violation

Table II lists the operating costs of DA markets and FMMs for both the FRP designs. Although, the proposed approach results in the higher DA operating costs in comparison to the other model (only 0.3% increase), it has much fewer expensive adjustments in the FMMs and has less real-time operating costs in the FMMs. It is evident from these results that the DA operating cost relatively has not been changed with respect to the corresponding average real-time operating costs for the proposed approach, while huge difference can be seen for the other approach. The average real-time operating cost of the general approach over all of the scenarios in the FMM with original bids is equal to $ 1692k, which is considerably higher than that of the proposed method, i.e., $ 1483k. On the other hand, the standard deviation of real-time operating costs of the proposed approach is effectively less than that of the general approach, which demonstrates the robustness of the proposed approach performance in response to the different realized net load scenarios. Finally, from evaluating the results for increased bid of FS units in Table II, it can be seen that the performance of the proposed approach is consistent with the non-increased bids results explained above. In Table III, five statistical measures are utilized to assess and compare the reliability extent in the FMMs that can be resulted from implementing two approaches. The results confirm that the proposed approach outperforms the general approach with respect to all violation statistical measures because the proposed FRP design considers the 15-min net load changes on the dispatch and commitment of the DA market to attain more adequate responsiveness from the flexible resources in RT (resource adequacy). Note that effectively decreasing the max violation is mainly useful when the ISO is interested in decreasing worst-case violation.

TABLE II. RESULTS FOR DA MARKET AND FMMS (FIRST TEST DAY)

| Approach | General FRP design | Proposed FRP design | General FRP design (with increased bid of FS units in FMMs) | Proposed FRP design (with increased bid of FS units in FMMs) |
|---|---|---|---|---|
| DA operating cost (K$) | 1420 | 1424 | 1420 | 1424 |
| Real-time operating costs in FMMs | | | | |
| Ave (K$) | 1692 | 1483 | 1693 | 1483 |
| Standard deviation (K$) | 478 | 184 | 478 | 183 |
| Max (K$) | 3977 | 2701 | 3979 | 2702 |

TABLE III. VIOLATION COMPARISON IN RT OPERATION (FIRST TEST DAY)

| Metric | General FRP design | Proposed FRP design |
|---|---|---|
| Average [violation] (MWh) | 6.6 | 1.4 |
| Standard deviation [violation] (MWh) | 11.9 | 4.6 |
| Σ violation (MWh) | 3298.8 | 702.3 |
| # Scenarios with violation | 204 | 73 |
| Max violation (MWh) | 63.6 | 31.8 |

The metrics presented in Table IV compare the 15-min commitments number of FS units that are turned on in the FMMs to follow the 15-min net load changes, in addition to the FS units that were previously committed from the DA market. Four metrics, which are indicative of the discrepancy between DA and RT operations, are presented in this table. It can be observed that the proposed model effectively reduces the need for the expensive FS units to be additionally committed in the RT market by turning on cheaper flexible units which are available in the DA market.

TABLE IV. RESULTS FOR INCREASED NUMBER OF 15-MIN COMMITMENTS OF FS UNITS IN RT OPERATION (FIRST TEST DAY)

| Metric | General FRP design | Proposed FRP design |
|---|---|---|
| Ave [Increased commitment of FS units] | 125 | 44 |
| Σ increased commitment of FS units | 62278 | 22083 |
| # Scenarios with increased commitment of FS units | 498 | 377 |
| Max increased commitment of FS units | 336 | 192 |

The proposed approach is also assessed on the net load scenarios from one additional test day in order to evaluate its robustness. Table V summarizes the corresponding results across 15-min net load scenarios for this day. Consistent with the first test day results, the proposed model outperforms the general approach by improving the reliability of system while also reducing the final operating costs in FMMs.

C. *Market implications of modeling FRP in DA*

Table VI summarizes DA energy revenue, FRP up revenue, and FRP down revenue for each approach. The DA FRP procurement aids in ensuring the DA resource adequacy to meet the net load changes in RT, which potentially leads to increase of the energy revenue in DA models with FRP requirements in comparison to the DA model without FRP requirements. Furthermore, it is evident that, as quantification of the DA FRP requirements becomes more accurate, the revenue of the generators for providing the ramp capabilities increases. Table VII presents the DA market settlement results for the different approaches. According to Table VII, the DA model with general FRP design has the highest generation revenue and generation rent, while the lowest value for these market settlements belongs to the DA model without FRP design. It is pertinent to note that the generation revenue is the result of summation over the energy revenue and FRP up and down revenues. Additionally, the lowest and highest congestion rents belong to DA model without FRP and DA model with general FRP design, respectively. Finally, in terms of load payment, the DA model with general FRP design has the highest value, and the DA model without FRP design has the lowest value.

TABLE V. RESULTS FOR DA MARKET AND FMMS (SECOND TEST DAY)

| Approach | General FRP design | Proposed FRP design |
|---|---|---|
| DA operating cost (K$) | 1647.8 | 1649.3 |
| Real-time operating costs in FMMs | | |
| Average (K$) | 1731.8 | 1665.5 |
| Standard deviation (K$) | 224.8 | 84.5 |
| Max (K$) | 3093.7 | 2661.0 |
| Violation Comparison in RT operation | | |
| Average [violation] (MWh) | 2.0 | 0.4 |
| Standard deviation [violation] (MWh) | 5.6 | 2.1 |
| Σ violation (MWh) | 980.7 | 177.8 |
| # Scenarios with violation | 90 | 27 |
| Max violation (MWh) | 36.0 | 25.0 |
| Increased number of 15-min commitments of FS units in RT operation | | |
| Ave [Increased commitment of FS units] | 91 | 33 |
| Σ increased commitment of FS units | 45590 | 16634 |
| # Scenarios with increased commitment of FS units | 489 | 386 |
| Max increased commitment of FS units | 269 | 146 |

TABLE VI. DA MARKET REVENUES COMPARISON

| Approach | Energy revenue (k$) | FRP up revenue (k$) | FRP down revenue (k$) |
|---|---|---|---|
| DA market without FRP | 1622.8 | 0 | 0 |
| DA market with general FRP design | 1644.7 | 5.6 | 0 |
| DA market with proposed FRP design | 1627.1 | 11.2 | 2.9 |

From the locational marginal prices (LMPs) stand view in the FMMs, in all scenarios and intervals, the general approach has 453 cases with price spikes, while the proposed approach has only 169 cases with price spikes. Table VIII shows distribution of these cases. It is clear that the proposed approach is able to considerably reduce the frequency of price spikes in the RT market, which potentially can prevent market inefficiency.

D. *Sensitivity Analyses*

Figures 7 and 8 illustrate the results of sensitivity analyses of the reliability and economic metrics with respect to VOLL and the percentage of DA net load uncertainty for the proposed FRP design in comparison to the general FRP design. These results show robustness of the proposed approach as the percentage of improvements of all metrics remains considerable.

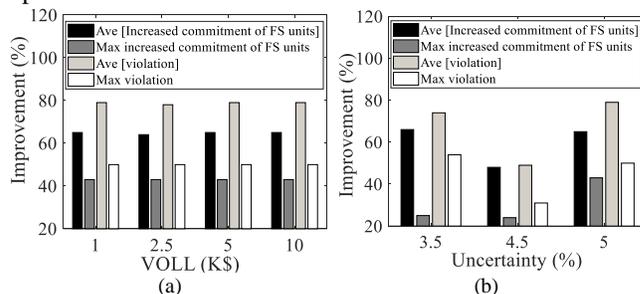

Fig. 7. Sensitivity analyses of violation and increased FS 15-min commitment with respect to: a) VOLL, and b) percentage of uncertainty.

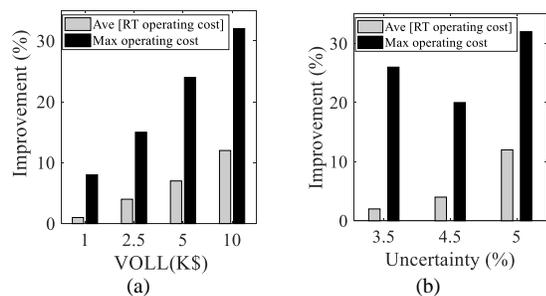

Fig. 8. Sensitivity analyses of RT operating costs with respect to: a) VOLL, and b) percentage of uncertainty.

TABLE VII. DA MARKET SETTLEMENTS COMPARISON

| Approach | Generation revenue (k$) | Load payment (k$) | Generation rent (k$) | Congestion rent (k$) |
|---|---|---|---|---|
| DA market without FRP | 1622.8 | 1622.8 | 220.7 | 0.0 |
| DA market with general FRP | 1650.3 | 1645.4 | 245.3 | 0.7 |
| DA market with proposed FRP | 1641.2 | 1627.5 | 233.4 | 0.3 |

TABLE VIII. DISTRIBUTION OF PRICE SPIKES AMONG CASES IN RT MARKET

| Approach | Number of cases |
|---|---|
| General approach only with price spike | 320 |
| Proposed approach only with price spike | 36 |
| Both approaches simultaneously have price spike | 133 |

## VI. CONCLUSION

The high penetration of the renewable resources is imposing new challenges (e.g., potential ramp capability shortage) to modern power systems by intensifying the uncertainty and variability in the system net load. The ramp capability shortage not only can jeopardize reliability, but also can cause inefficiencies in the RT markets. Energy markets are evolving to overcome such challenges by implementing new ancillary service products called FRP. There have been proposals to implement the FRP into the DA markets, e.g. CAISO DA market, in order to ensure resource adequacy to respond to net load variability and uncertainty in the next market processes such as the FMM and the RT economic dispatch. The DA formulation with FRP proposed by existing work is based on the hourly ramping requirements with the aim to enhance ramp capabilities in the shorter scheduling granularity, e.g., 15 min. However, it ignores the impacts of steep 15-min net load changes while scheduling DA FRP awards. To address this challenge, this paper proposes a novel DA FRP design to preposition and commit the ramp-responsive resources to better respond to the RT ramping needs. The proposed DA FRP formulation are designed in such a way that (i) the hourly net load variability and uncertainty are taken into account, and (ii) impacts of the 15-min net load variability and uncertainty are captured in the awarded DA FRP. The proposed model is more adaptive to the RT condition through enhancing the quantity allocation of FRP by making relatively small market changes compared to the general FRP design. The proposed approach leads to less expected final operating cost in the FMMs, higher reliability as the power system gets close to RT operation, and less discrepancy between DA and FMMs decisions. Furthermore, corresponding incentive policies were derived based on the duality theory for the proposed FRP design, and then a comprehensive evaluation of the DA market implications of new reformulation was presented and discussed.